\documentclass[reprint,aps,prd,superscriptaddress,showkeys,showpacs]{revtex4-1}
\usepackage{epsfig,amsmath,natbib}
\usepackage{aas_macros}
\usepackage{amssymb}
\usepackage{amsmath}
\usepackage{dsfont}
\usepackage{hyperref}
\usepackage{color}
\usepackage{pbox}
\usepackage{enumerate}
\usepackage{multirow}

\hypersetup{
	colorlinks=true,
	linkcolor=blue,
	urlcolor=blue,
	citecolor=blue
}
\newcommand\lsim{\mathrel{\rlap{\lower4pt\hbox{\hskip0pt$\sim$}}
        \raise1pt\hbox{$<$}}}
\newcommand\gsim{\mathrel{\rlap{\lower4pt\hbox{\hskip0pt$\sim$}}
        \raise1pt\hbox{$>$}}}

\newcommand\DNN{\operatorname{DNN}}
\newcommand\lin{\operatorname{lin}}

\usepackage[dvipsnames]{xcolor}

\begin{document}

\title{Interpreting deep learning models for weak lensing}

\author{Jos\'e Manuel Zorrilla Matilla}
\email{jzorrilla@astro.columbia.edu}
\affiliation{Department of Astronomy, Columbia University, New York, NY 10027, USA}
\author{Manasi Sharma}
\affiliation{Department of Computer Science, Columbia University, New York, NY 10027, USA}
\author{Daniel Hsu}
\affiliation{Department of Computer Science, Columbia University, New York, NY 10027, USA}
\author{Zolt\'an Haiman}
\affiliation{Department of Astronomy, Columbia University, New York, NY 10027, USA}

\date{\today}

\begin{abstract}

Deep Neural Networks (DNNs) are powerful algorithms that have been proven capable of extracting non-Gaussian information from weak lensing (WL) data sets. Understanding which features in the data determine the output of these nested, non-linear algorithms is an important but challenging task. We analyze a DNN that has been found in previous work to accurately recover cosmological parameters in simulated maps of the WL convergence ($\kappa$). We derive constraints on the cosmological parameter pair $(\Omega_m,\sigma_8)$ from a combination of three commonly used WL statistics (power spectrum, lensing peaks, and Minkowski functionals), using ray-traced 
simulated $\kappa$ maps.  We show that the network can improve the inferred parameter constraints relative to this combination by $20\%$ 
even in the presence of realistic levels of shape noise. 
We apply a series of well established saliency methods to interpret the DNN and find that the most relevant pixels are those with extreme $\kappa$ values. For noiseless maps, regions with negative $\kappa$ account for $86-69\%$ of the attribution of the DNN output, defined as the square of the saliency in input space.
In the presence of shape nose, the attribution concentrates in high convergence regions, with $36-68\%$ of the attribution in regions with $\kappa > 3 \sigma_{\kappa}$.
\end{abstract}
\keywords{}
\pacs{}
\maketitle

\section{Introduction}\label{cnninterp.introduction} 
The perturbed trajectories of photons propagating through an inhomogeneous universe result in (de)magnified and distorted images of background galaxies. The effect is in general very small, but can be detected through measurements over a large ensemble of galaxies. This weak gravitational lensing (WL) can be used to reconstruct the matter density field between us, observers, and the lensed background galaxies~\citep{Bartelmann2001, Kilbinger2015}. The statistical properties of this field, and its evolution, can be used to test the standard $\Lambda$ + cold dark matter ($\Lambda$CDM) cosmological model. WL measurements are particularly sensitive to two of the defining parameters of $\Lambda$CDM: the mean matter density of the universe, $\Omega_m$, and the amplitude of the initial perturbations that acted as seeds for the growth of structure, which can be measured in the local universe as $\sigma_8$. Recent estimates for those parameters using WL measurements hint at a possible tension with values inferred from observations of the cosmic microwave background~\citep{Asgari2020, Abbott2019, Hikage2019,LuHaiman2019}, strengthening the case for more precise measurements.

Upcoming galaxy surveys, such as the Vera Rubin Observatory (VRO) Legacy Survey of Space and Time (LSST~\cite{LSST2009}), the Euclid space mission~\citep{Euclid2010}, and the Wide Field Infrared Survey Telescope (WFIRST~\cite{WFIRST2015}), will provide those measurements in the near future. It is crucial to extract and use optimally all the cosmological information encoded in the measurements obtained by these experiments. Some of the data products will be mass maps of unprecedented angular resolution, with the projected matter density (convergence, or $\kappa$) up to a certain redshift.

The standard method to estimate cosmological parameters from such maps, within a Bayesian framework, is to compress the information content of all the pixels into a single data vector, or summary statistic, for which a likelihood can be computed or sampled using simulations. These summary statistics can be physically motivated (e.g. the use of the power spectrum, supported by the independent evolution of different Fourier modes in the linear regime of the growth of structure), or aim to describe the map morphology (e.g. the total length of isocontours, or second Minkowski functional of a 2D field). The choice of any particular statistic will, however, generically entail a loss of information. 

An alternative approach is to bypass the design of summary statistics, and use a data-driven algorithm to map directly the pixels in a map onto the parameters of interest. Recent attempts of doing so with deep neural networks (DNNs) have shown that these algorithms can provide competitive parameter constraints. This has been demonstrated not only for simulated data~\citep{Gupta2018, Fluri2018, Ribli2019b}, but also for WL survey data~\citep{Fluri2019}. While DNNs are capable of learning complex non-linear relationships between data and the parameters that control the generative models behind the data, they are notoriously difficult to interpret. This is due the large number of fitted parameters (weights and biases) involved, and the depth of the many layers of non-linearities that comprise a DNN.

Previous studies have attempted to understand DNN models trained on WL data. The feature maps output by a model's intermediate layers have been found difficult to interpret~\cite{Schlemzle2017}, and the same applies to intermediate convolution filters~\cite{Merten2019}. The analysis of the convolution filters on the first DNN layer has proven more fruitful, with at least one example~\citep{Ribli2019a} of filters that could be interpreted and used to design a new powerful summary statistic (the distribution of the radial profiles of local maxima, or peaks, in the maps). The limitation of the analysis of feature maps and learned filters to interpret DNNs is that they do not take into account the impact of the identified filters on the model's output, which can be complicated by non-linear interactions with other components of the DNN,  and as a result cannot connect features in the input data space to the output of the networks. 

The aim of the present study is to interpret a high-performing DNN trained on WL data~\citep{Ribli2019b} using state-of-the-art attribution methods from the field of image classification. These so-called saliency methods have been developed to understand the output of DNNs by providing an attribution or importance metric for each individual pixel of a given input datum. This is a fast-evolving field, and many such methods have been proposed~\citep{Simonyan2013, Springenberg2014, Bach2015, Montavon2015, Shrikumar2016, Zintgraf2017, Smilkov2017, Fong2017}. For an in-depth review of the sub-field of explainable DNN models, we refer the reader to~\cite{Samek2019}.

This paper is organized as follows. First, we describe the DNN we study and the data used to train it, in \S~\ref{cnninterp.datamodel}.  In \S~\ref{cnninterp.performance}, we then assess the performance of the model relative to a {\em combination} of summary statistics typically used to analyze WL data to confirm that the model processes information not accessible to these statistics. In \S~\ref{cnninterp.attributions} we evaluate a series of attribution methods for DNNs proposed in the literature, and select the more appropriate ones for our combination of model and data. We use the selected methods to study features in input space with the largest impact on the DNN's output.  Finally, we discuss our results and summarize our conclusions in~\S~\ref{cnninterp.discussandconclude}.

\section{Model and data} \label{cnninterp.datamodel} 
The DNN model analyzed in this study is the one developed in \cite{Ribli2019b}. It is an architecture that combines 2D convolutional layers (18) and average pooling layers (6) to map inputs consisting of simulated WL converge ($\kappa$) maps into two parameters of interests. Each convolutional layer is followed by batch normalization, except for the last one. All activation functions are rectified linear units (ReLUs), and the network was trained using stochastic gradient descent and a mean absolute error loss function.  We will often refer to this specific DNN, including its architecture and learned parameters, as simply the ``model''.

The data set used for training and evaluation of the model is a suite of simulated $\kappa$ maps. These maps are built following the trajectories of bundles of light rays along simulated past light cones up to the redshift where the lensed galaxies are assumed to lie. Each past light cone is assimilated to a discrete set of lensing planes responsible for the light rays' deflections due to gravitational lensing (so-called "multi-plane algorithm"). The lensing planes are computed from the output of dark-matter-only N-body simulations, which mimic the time evolution of the matter density field, given a set of initial conditions. The resulting convergence maps can be interpreted (to first order) as the anisotropies in the projected matter density field weighted by an appropriate lensing efficiency kernel. This data set has been used in past studies of deep learning applied to weak lensing~\citep{Gupta2018, Ribli2019a, Ribli2019b}, and we refer the reader to these references for a detailed description of the simulation pipeline. The full suite consists in synthetic convergence maps for 101 different cosmologies, each defined by a distinct pair of parameters $\{\Omega_m, \sigma_8\}$, corresponding to the mean matter density of the universe (in units of its critical density) and the amplitude of the initial perturbations normalized in the local universe. For each cosmology, 512 independent $3.5\times3.5\,{\rm deg^2}$ maps were generated.

In this first academic study, we focus on the interpretation of a model trained on two sets of maps. Both sets assume that all the lensed galaxies lie at the same redshift of $z=1$. In both sets, maps consist of $512\times512$ pixels---each having a linear angular size of $0.41\,{\rm arcmin}$--- and are smoothed using a Gaussian filter with $\sigma=1\,{\rm arcmin}$.  The first set corresponds to noiseless data, and the second includes galaxy shape noise, modeled as a Gaussian white noise with zero mean and a standard deviation, $\sigma_n$, determined by the galaxies' intrinsic ellipticity, $\sigma_{\epsilon}=0.4$, the survey's galaxy number density, $n_g=30\,{\rm arcmin^{-2}}$, and the pixel area, $A_{pix}$:
\begin{equation}
    \sigma_n = \frac{\sigma_{\epsilon}}{\sqrt{2n_g A_{pix}}}.
\end{equation}

The values considered for the galaxies' intrinsic ellipticity and density are comparable values expected for future experiments. For example, an ellipticity of 0.4 corresponds to that from the Subaru Hyper Suprime-Cam Survey (HSC~\cite{Mandelbaum2018}), and a density of 30 arcmin$^{-2}$ will be achievable by WFIRST up to a redshift of $z=1$~\cite{WFIRST2015}.

The data was split into a training and a test set, encompassing $70\%$ and $30\%$ of the maps, respectively. The network was trained so that its output predicts  $\{\Omega_m, \sigma_8\}$, in units of the standard deviation of those parameters in the data set.

\section{Network performance relative to alternative statistics}\label{cnninterp.performance} 
Past studies have shown that neural networks offer a discriminating power between cosmological models competitive with alternative statistics, such as the power spectrum~\citep{Gupta2018, Fluri2018, Fluri2019}, lensing peaks~\citep{Gupta2018, Ribli2019b}, skewness or kurtosis~\citep{Schlemzle2017} and a combination of the power spectrum, lensing peaks and Minkowski functionals (MFs)~\citep{Merten2019}.

We begin by comparing the performance of the DNN model to a combination of the power spectrum, lensing peaks and Minkowski functionals, to assess if the DNN exploits information not accessible through these well-explored summary statistics of weak lensing fields. 

The power spectrum is the Fourier transform of the two-point correlation function. It fully characterizes Gaussian random fields, such as the matter density field after recombination, and is a commonly-used statistic in cosmology. However, as gravitational collapse induces non-Gaussianities in the matter density field, additional statistics are needed to extract all the information encoded in lensing data sets. The $\kappa$ power spectrum was measured on the simulated data set as an azimuthal average of the squared Fourier transforms of the maps. For the galaxy distribution considered, with all lensed galaxies at $z=1$, and a flat universe, the measured power spectrum corresponds to the (auto) convergence power spectrum $P_{\kappa}$, which can be related to the 3D matter power spectrum $P_{\delta}$ using the Limber approximation:

\begin{equation}
\resizebox{.43 \textwidth}{!}
{
    $P_{\kappa}\left(\ell\right)=\left(\frac{3H_0^2\Omega_m}{2c^2}\right)^2 \int_0^{\chi_s} \frac{d\chi}{a^2\left(\chi\right)}\left(1-\frac{\chi}{\chi_s}\right)^2 P_{\delta}\left(\frac{\ell}{\chi},\chi\right)$
}
\end{equation}

\noindent where $\ell$ is the angular wavenumber, $H_0$ the Hubble constant, $c$ the speed of light, $\chi$ the comoving radial distance, and $\chi_s$ the comoving radial distance to the lensed galaxies' redshift.

Counts of local maxima as a function of their height, or ``lensing peaks'', is a statistic that is simple to measure and has been shown to improve constraints derived from the power spectrum alone by up to a factor of $\approx$ two~\citep{Kratochvil2010, Dietrich2010}. Lensing peaks have also been successfully used to analyze survey data and to improve parameter constraints~\citep{Liu2015, LiuPan2015, Kacprzak2016, Martinet2018}. On our simulated maps, peaks correspond to pixels whose $\kappa$ value exceeds that of their six neighboring pixels.

Three different Minkowski functionals can be defined for two-dimensional fields~\citep{Mecke1994, Schmalzing1996} by performing measurements over excursion sets defined by the points whose value exceeds a given threshold. The first one, V$_0$, measures the area of the excursion set, the second V$_1$, the total length of its boundary, and the third, V$_2$, its genus. They were measured over the simulate maps from the gradients $\kappa_{x, y}$ (estimated by finite difference):

\begin{widetext}
\begin{gather}
    V_0\left(\kappa_0\right) = \frac{1}{A}\int_A d\mathbf{\theta}\Theta\left(\kappa\left(\mathbf{\theta}\right)-\kappa_0\right) \nonumber \\
    V_1\left(\kappa_0\right) = \frac{1}{4A} \int_A d\mathbf{\theta} \delta_D\left(\kappa\left(\mathbf{\theta}\right)-\kappa_0\right) \sqrt{\kappa_x^2 + \kappa_y^2}\\
    V_2\left(\kappa_0\right) =\frac{1}{2\pi A} \int_A d\mathbf{\theta} \delta_D\left(\kappa\left(\mathbf{\theta}\right)-\kappa_0\right)\frac{2\kappa_x\kappa_y\kappa_{xy}-\kappa_x^2\kappa_{yy}-\kappa_y^2\kappa_{xx}}{\kappa_x^2 + \kappa_y^2} \nonumber 
\end{gather}
\end{widetext}

\noindent where A is the area of the map, $\Theta$ the Heaviside function and $\delta_D$ the Dirac delta function. The Minkowski functionals have been shown to improve constraints derived from the power spectrum by a factor of up to $\approx 2-3$~\citep{Munshi2012, Petri2015, Ling2015, Marques2019}.

We combined the power spectrum, lensing peaks and the three Minkowski functionals into a single data vector, and estimated the constraints on the parameters $\{\Omega_m, \sigma_8\}$ assuming the Gaussian likelihood:
\begin{equation}\label{cnninterp.eq.likelihood}
P\left(\mathbf{s}|\boldsymbol{\theta}\right) \propto \exp\left\{-\frac{1}{2}\left[\mathbf{s} -  \bar{\mathbf{s}}\left(\boldsymbol{\theta}\right)\right] \widehat{{\rm C}^{-1}} \left[\mathbf{s} -  \bar{\mathbf{s}}\left(\boldsymbol{\theta}\right)^T\right] \right\},
\end{equation}
where $\mathbf{s}$ is the measured data vector, $\bar{\mathbf{s}}\left(\boldsymbol{\theta}\right)$ the expected value of the data vector in a cosmology defined by the parameter set $\boldsymbol{\theta}$, and $\widehat{{\rm C}^{-1}}$ the estimated precision matrix, which we evaluate at the single (fiducial) cosmology defined by $\Omega_m=0.260$ and $\sigma_8=0.8$ (so there is no need to consider the pre-factor with the covariance determinant).

Each statistic was measured in 20 bins. For the power spectrum, we considered uniformly spaced bins in logarithmic space, with spherical harmonic index between $\ell \in \left[100, 15000\right]$. For the lensing peaks and Minkowski functionals, we used uniformly spaced bins in linear space, between $\kappa \in \left[-0.0235, 0.0704\right]$ for the noiseless data, and $\kappa \in \left[-0.05, 0.10\right]$ for the data with shape noise, corresponding to $\left[-2,6\right]$ and $\left[-2.7,5.3\right]$ in units of the measured r.m.s.\ of the $\kappa$ field for the fiducial cosmology, respectively.
The expected data vector at a point in parameter space not present in the simulation suite is computed using an emulator. The emulator is a 2D Clough-Tocher interpolator (as implemented in the Python SciPy library~\cite{SciPy2020S}), fitted on the mean values of the data vector measured on the test data set. We analyzed the impact on the parameter inference of possible interpolation errors due to the discrete sampling of the parameter space by our simulations, and found it negligible (see Appendix~\ref{cnninterp.appendix_A}).

The covariance matrix of the combined statistics data vector was estimated from a new suite of simulations in the fiducial cosmology, consisting of 120,000
independent $\kappa$ maps ray-traced through the outputs of 100 new, fully independent N-body realizations of the underlying matter density field. We verified that this number of maps suffices via cross-validation  ---the credible contours for $\Omega_m$ and $\sigma_8$ from the combined statistics measured on three independent sub-samples of the data, each consisting of 40,000 maps, are indistinguishable from each other, indicating numerical convergence. The bias in the estimation of the inverse covariance is accounted for by applying the correction factor $\frac{N-d-2}{N-1}$ (where $N=120,000$ is the number of measurements and $d=102$ is the dimension of the data vector~\cite{Hartlap2007}). We note that, when using estimated covariance matrices, the likelihood is not Gaussian anymore, but an adapted version of the multivariate t-distribution~\citep{Sellentin2016}. An additional correction may be needed when the data-vector is estimated from noisy simulations~\citep{Jeffrey2019}. We did not consider these corrections, which we deem small due to the large number of simulated maps available relative to the length of the data vector for which the covariance is estimated.

We treated the cosmological parameters $(\Omega_m,\sigma_8)$ output by the DNN as just another summary statistic that can be added to the data vector, increasing its size from 100 to 102. This allowed us a uniform treatment of the DNN output and the other summary statistics within the framework of a Gaussian likelihood (for a test of Gaussianity of a DNN output, see~\cite{Gupta2018}). Neither the means nor the covariance estimates used any of the $\kappa$ maps present in the network's training data set.

\begin{figure*}
    \centering
    \includegraphics[width=\linewidth]{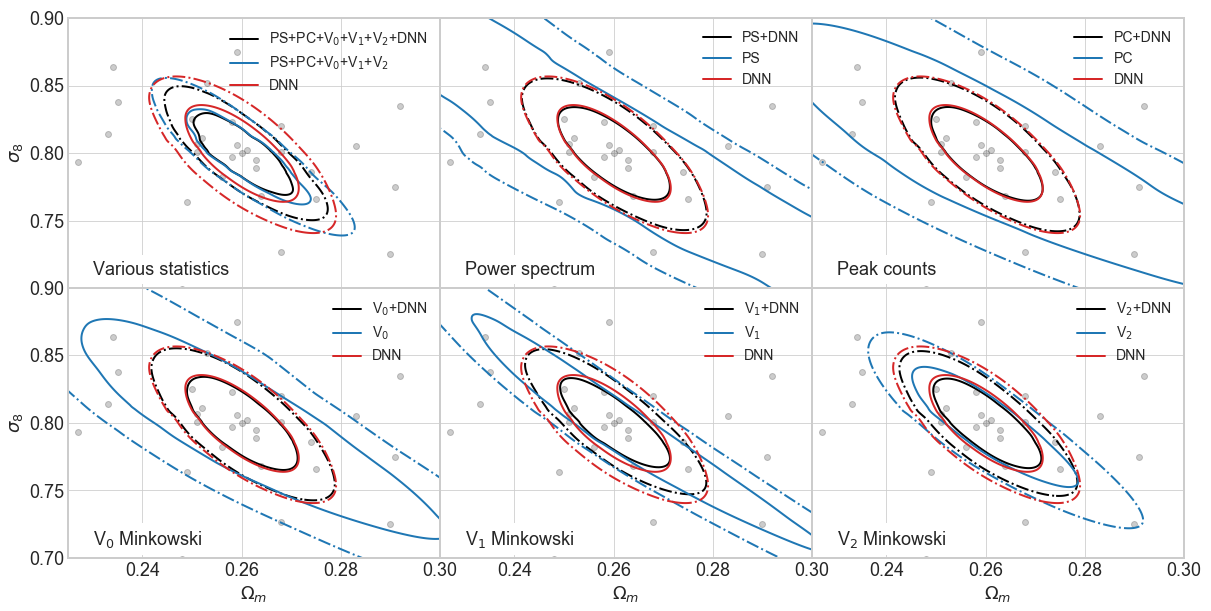}
    \caption{Credible contours derived for $\Omega_m$ and $\sigma_8$. Each panel shows the comparison between the constraints derived from the DNN (in red) from an alternative statistic (in blue), and the combination of the DNN and the statistic (in black). Solid lines enclose 68\% of the likelihood, and dot-dashed lines 95\%. Upper row, from left to right: comparison between the DNN and a combination of statistics, the power spectrum (PS), and lensing peak counts (PC). Lower row, from left to right: comparison between the DNN and the three Minkowski functionals, V$_0$, V$_1$, and V$_2$. The grey dots indicate the points in parameter space for which simulations were available.
    }
    \label{fig:cnn.interp.performance}
\end{figure*}

\begin{figure*}
    \centering
    \includegraphics[width=\linewidth]{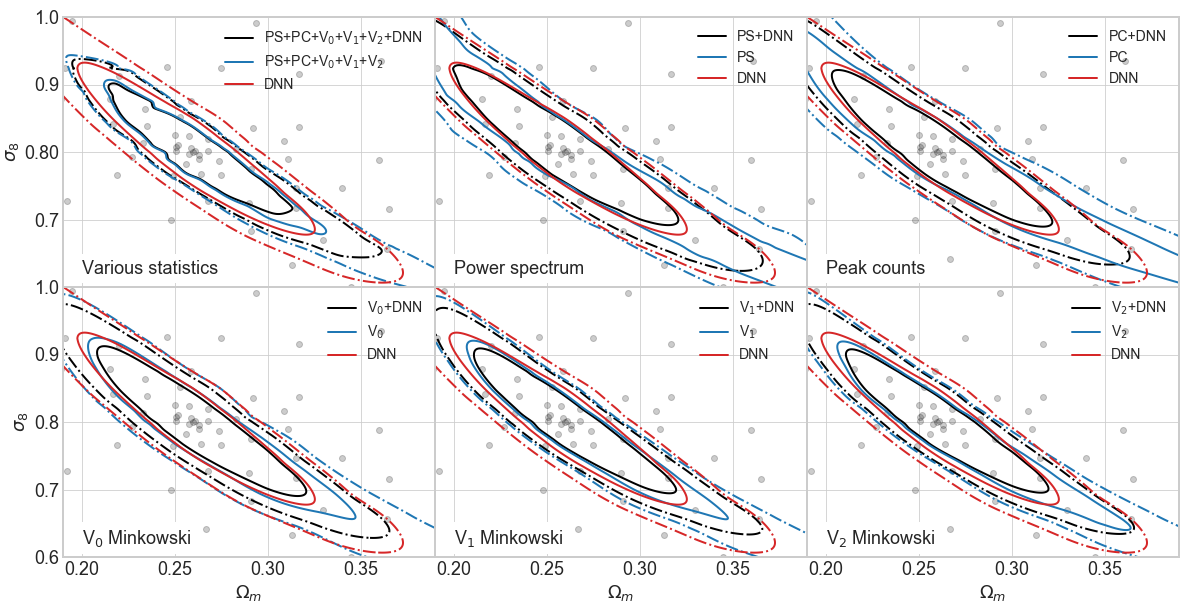}
    \caption{Same as Fig.~\ref{fig:cnn.interp.performance} for statistics measured on, and the DNN trained on, maps that include the effect of shape noise for a galaxy density of $n_g=30\,{\rm arcmin^{-2}}$ and intrinsic ellipticity of $\sigma_{\epsilon}=0.4$.
    }
    \label{fig:cnn.interp.performance.noisy}
\end{figure*}

In Fig.~\ref{fig:cnn.interp.performance}, we show the credible contours for $\Omega_m$ and $\sigma_8$ that can be derived from noiseless maps using summary statistics, the DNN, or a combination of both. The constraints are displayed separately for each individual statistic, and a combination of the power spectrum, lensing peaks, and Minkowski functionals. The same contours in the presence of shape noise are shown in Fig.~\ref{fig:cnn.interp.performance.noisy}. The percentage change in the area of the credible contours achieved when the output of the DNN is incorporated, defined as $\Delta {\rm Area} = 100\left(\frac{Area_{\rm w/\,DNN}}{Area_{\rm w/o\,DNN}}-1\right)$, is reported in Table~\ref{table:cnn.interp.performance}. Since the focus of this analysis is the relative performance of different summaries extracted from the data, rather than estimating the expected constraining power for a given survey definition, none of the credible contours in this manuscript are scaled to match a particular sky area; they correspond to the simulated maps' $3.5\times3.5\,\rm deg^2$ area.

The improvement relative to the power spectrum and lensing peaks is very substantial, as has been shown in past studies~\citep{Gupta2018,Fluri2018,Ribli2019b,Fluri2019}. The improvement relative to the Minkowski functionals is more modest, in particular compared with V$_2$, which is by far the most constraining when measured on noiseless $\kappa$ maps. In the presence of noise, V$_2$ degrades more than other Minkowski functionals, and V$_1$ turns into the most constraining non-Gaussian statistic. V$_1$ and V$_2$ are particularly sensitive to void regions. The fact that they degrade more than other statistics in relative terms in the presence of shape noise may point to a shift in the amount of cosmological information that can be recovered from them.

When combining all the statistics together, the addition of the DNN predictions manages to reduce the area of the credible contours by $\approx20\%$, in both the noiseless and noisy cases. While the DNN does not tighten the contours by a large factor, the difference is significant, implying that the DNN can extract information in the maps that is not accessible to the alternative statistics. 

\begin{table}
\begin{center}
\begin{tabular}{c c c c c c c c}
\hline
\hline
&& \multicolumn{6}{c}{Change in credible contour area [\%]}\\
\cline{3-8}
Dataset & Credibility & PS & PC & V$_0$ & V$_1$ & V$_2$ & All\\
\hline
\multirow{2}{*}{Noiseless} & 68\% & -93 & -89 & -81 & -68 & -38 & -19 \\
              & 95\% & -93 & -89 & -80 & -69 & -38 & -19 \\
\hline
\multirow{2}{*}{Noisy}     & 68\% & -70 & -54 & -33 & -25 & -30 & -18 \\
              & 95\% & -64 & -55 & -37 & -30 & -32 & -22 \\
\hline
\hline
\end{tabular}
\caption{Percentage change in the area of credible contours derived from different statistics when they are combined with the output from the DNN (see Fig.~\ref{fig:cnn.interp.performance} for a graphical representation of those contours). The statistic used for the right-most column (labelled ``All'') is a combination of all the statistics in the other columns: power spectrum (PS), lensing peaks (PC), and Minkowski functionals (V$_0$, V$_1$, V$_2$). The change in area is defined as $\Delta {\rm Area} = 100\left(\frac{\rm Area_{\rm w/\,DNN}}{\rm Area_{\rm w/o\,DNN}}-1\right)$.}\label{table:cnn.interp.performance}
\end{center}
\end{table}

Since the DNN does not improve constraints from the combination of the summary statistics by a large factor, it is worth verifying that the network is not essentially learning the statistics. First, we looked at the Pearson's correlation coefficient between the DNN's output and the measured statistics, conditioned to the true value of the parameters $\{\Omega_m, \sigma_8\}$, that is, we computed the correlation for each cosmology in the test data set, and took the average of the values for the 101 cosmologies (the cosmology dependence of the correlations is weak).

\begin{figure*}
    \centering
    \includegraphics[width=\linewidth]{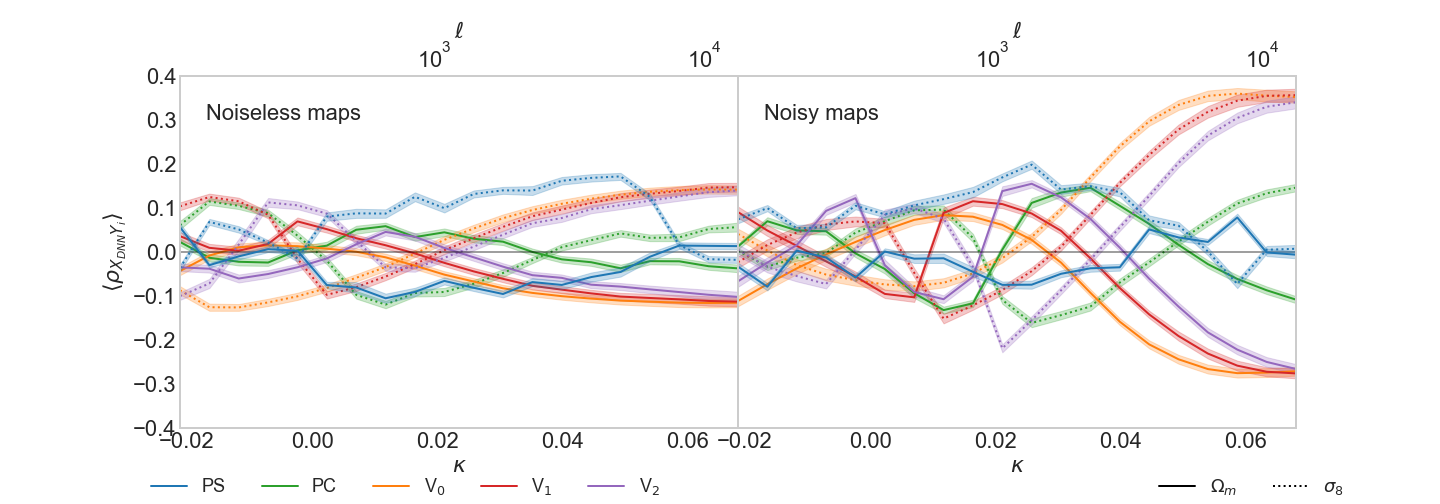}
    \caption{Pearson's correlation coefficient (averaged over the 101 cosmologies) between the DNN predictions for $\Omega_m$ (solid lines) and $\sigma_8$ (dotted lines), and the measured statistics. For the power spectrum (PS, blue), the bins correspond to different multipoles ($\ell$; see upper scale), and for the other statistics, values of $\kappa$ (lower scale). Shaded regions represent uncertainties in the plotted means, estimated as $\sigma / \sqrt{101}$, with $\sigma$ the standard deviation of the correlation coefficient over the 101 cosmologies. The left panel corresponds to correlations from noiseless maps, and the right panel to correlations in the presence of shape noise.}
    \label{fig:cnn.interp.statcorr}
\end{figure*}

Fig.~\ref{fig:cnn.interp.statcorr} shows the average correlation coefficients as a function of $\ell$ bin for the power spectrum (PS) and $\kappa$ bin for the other statistics (computed from the test data set). None of the correlations is particularly high (all of the coefficients are below 0.2 in absolute value for the noiseless case, and 0.4 in the presence of shape noise). The correlations for $\Omega_m$ and $\sigma_8$ tend to have opposite sign, indicating that the DNN learned the degeneracy between the two parameters. Besides, the qualitative change in the correlations as a function of binning follows expectations. For instance, higher $\sigma_8$ is positively correlated with higher power spectrum, yielding a larger $\sigma_{\kappa}^2$, with results in a lower central peak and fatter tails for the lensing peak distribution.

In the presence of shape noise, the correlation between the predictions of the DNN and the high-significance peaks (and Minkowski functionals measured over high-$\kappa$ thresholds) increases. This suggest that high-$\kappa$ regions become more important for the network in the presence of shape noise, as details in the low convergence regions of the maps are dominated by noise. The analysis of the network's sensitivity to its input, presented in \S~\ref{cnninterp.attributionmapping} below, confirms this trend.

It is also straightforward to show that the network's output cannot be reproduced by a linear combination of the summary statistics either. For each cosmology, we fit a linear model of the statistics for the output of the DNN:
\begin{equation}
\mathbf{y} = \mathbf{y_0} + \mathbf{W} \mathbf{s},
\end{equation}
\noindent{%
where $\mathbf{y}$ is the 2-dimensional output of the network, $\mathbf{s}$ a vector of length 100 with all five measured statistics stacked, and $\mathbf{y_0}$ (vector of length 2) and $\mathbf{W}$ (matrix of dimension $2\times100$) are determined to minimize the error of the model on the test data set.
}

We then computed the correlation coefficients between the actual output of the network $\{\Omega_m^{\rm DNN}, \sigma_8^{\rm DNN}\}$ and the output predicted by the linear model $\{\Omega_m^{\rm lin}, \sigma_8^{\rm lin}\}$, and average the correlations for the 101 cosmologies (the correlations do not exhibit any clear cosmology dependence).

The resulting average correlations go from weak ($\approx \pm 0.25$) to moderate ($\approx \pm 0.5$), with a maximum value of 0.62. For noiseless data: 
\begin{equation*}
    \begin{bmatrix}
    \rho\left(\Omega_m^{\DNN},\Omega_m^{\lin}\right) &
    \rho\left(\Omega_m^{\DNN},\sigma_8^{\lin}\right) \\
    \rho\left(\sigma_8^{\DNN},\Omega_m^{\lin}\right) &
    \rho\left(\sigma_8^{\DNN},\sigma_8^{\lin}\right)
    \end{bmatrix}
    =
    \begin{bmatrix}
    0.34 & -0.26 \\
    -0.24 & 0.45
    \end{bmatrix},
\end{equation*}

\noindent and in the presence of shape noise:

\begin{equation*}
    \begin{bmatrix}
    \rho\left(\Omega_m^{\DNN},\Omega_m^{\lin}\right) &
    \rho\left(\Omega_m^{\DNN},\sigma_8^{\lin}\right) \\
    \rho\left(\sigma_8^{\DNN},\Omega_m^{\lin}\right) &
    \rho\left(\sigma_8^{\DNN},\sigma_8^{\lin}\right)
    \end{bmatrix}
    =
    \begin{bmatrix}
    0.60 & -0.56 \\
    -0.56 & 0.62
    \end{bmatrix}
\end{equation*}

Given that (i) the DNN seems to access additional information and (ii) its outputs do not correlate highly with the summary statistics, or with their best-fit linear combination, we proceeded to look at the structure of the DNN to interpret its outputs.

\section{Interpreting DNNs with saliency methods} \label{cnninterp.attributions} 

DNNs can be interpreted as non-linear mappings from an input space of dimension $d$ (for this study, $512\times512$) to a space of dimension $n$ (for this study, two, the number of parameters of interest), $\mathcal{M}:\mathbb{R}^d\rightarrow \mathbb{R}^n$. Saliency methods map the input space into a space of the same dimension, $\mathcal{S}:\mathbb{R}^d\rightarrow \mathbb{R}^d$, so that the image of a given pixel, $\mathcal{S}\left(x_i\right)$ is representative of the importance of that pixel for a given output neuron, $\mathcal{M}\left(x_i\right)$.

We analyzed several established methods that are well-defined for network architectures utilizing rectified linear units (ReLUs~\cite{Nair2010}), and do not require re-training the model under study. These methods fall into two broad categories. The first category of saliency methods evaluates the effect of small perturbations of the input on the output. These methods rely on the the gradient of the DNN's output w.r.t.\ its input, which can be computed efficiently through a method called back-propagation---the iterative calculation of the gradient, layer-by-layer from the network's output to its input, avoiding redundant terms from the na\"{i}ve application of the chain rule~\citep{Kelley1960, Rumelhart1986}. We selected two gradient-based methods whose interpretation for linear models is straightforward:

\begin{itemize}

    \item \textbf{Gradient:} computes the gradient of the output neurons w.r.t.\ the values of the input pixels, $\mathcal{S}\left(x\right)=\frac{\partial \mathcal{M}}{\partial x}$. This measures the sensitivity of the output to the input, and for a linear model is equivalent to the regression coefficients.
    
    \item \textbf{Input$\times$gradient:} computes the element-wise product of the input and the gradient of the output w.r.t.\ the input pixels, $\mathcal{S}\left(x\right)=x \odot \frac{\partial \mathcal{M}}{\partial x}$. For a linear model, it measures the contribution of the pixel to the output.
    
\end{itemize}

Other gradient-based methods exist, such as {\it Smoothgrad}~\citep{Smilkov2017} or {\it Integrated gradients}~\citep{Sundararajan2017}, but we did not study these methods due to their significantly higher computational cost. We inspected their effect on a small subset of input maps, and the results were qualitatively very similar to those of the {\it Gradient} and {\it Input}$\times${\it gradient} methods.

The second category of saliency methods tries to distribute the network's output among the neurons of the second-to-last layer. The amount allocated to each neuron, interpreted as a relevance measure, is propagated iteratively through the network, back to the input space. We selected the following propagation-based methods:

\begin{itemize}

    \item \textbf{Guided backpropagation:} masks out negative gradients and negative activations when back-propagating the gradient of the output w.r.t.\ the input~\cite{Springenberg2014}. 
    
    \item \textbf{Deconvnet:} uses a deconvolution network~\cite{Zeiler2011}, $\mathcal{M}^{-1}$, built on top of the DNN architecture. To compute the saliency map corresponding to the input $x$, the feature maps $\{f^i\}$ for each layer $i$ in the model $\mathcal{M}$, are fed as inputs to the deconvolution network's layers. At each stage of the propagation through $\mathcal{M}^{-1}$, intermediate representations are unpooled, rectified, and filtered, until pixel space is reached~\cite{Zeiler2013}.
    
    \item \textbf{Deep Taylor decomposition:} distributes the relevance of neurons among its preceding layer by approximating the layer's function with a first order Taylor expansion~\cite{Montavon2015}. 
    
    \item \textbf{Layer-wise relevance propagation (LRP):} distributes the relevance of neurons among its preceding layer taking into consideration the weights that define the layer. We considered two different rules that are common in the literature.
    The first one, LRP-$\epsilon$ uses as rule to propagate the relevance: $R_i=\sum_j\frac{a_i w_{ij}}{\epsilon + \sum_i a_i w_{ij}}R_j$, where $a_i$ is the activation of neuron $i$, $w_{ij}$ the weight connecting neuron $j$ to neuron $i$, the relevances $R$ are the layer's output, and $\epsilon$ absorbs weak or contradictory contributions to the activations. For ReLU-based networks~\cite{Ancona2017}, $\epsilon=0$ renders this method equivalent to {\it Intput}$\times${\it gradient}. We chose $\epsilon=10^{-3}$, for larger values resulted in saliency maps indistinguishable from random noise. 
    The second rule, LRP-$\alpha\beta$, propagates the relevance according to: $R_i=\sum_j\left(\alpha \frac{\left(a_iw_{ij}\right)^+}{\sum_i \left(a_iw_{ij}\right)^+}- \beta \frac{\left(a_iw_{ij}\right)^-}{\sum_i \left(a_iw_{ij}\right)^-}\right)R_j$, where $()^+$ and $()^-$ refer to positive and negative contributions. We used $\alpha=1$ and $\beta=1$, a popular choice that renders this method equivalent to the {\it Excitation Backprop} method~\citep{Zhang2016}. We also validated that our results do not change qualitatively when the parameters $\alpha$ and $\beta$ are modified slightly.
\end{itemize}

We applied the same method to all of the layers in the DNN under study.

\subsection{Method comparison and selection}\label{cnninterp.attributions.choice} 

\begin{figure*}
    \centering
    \includegraphics[width=0.65\linewidth]{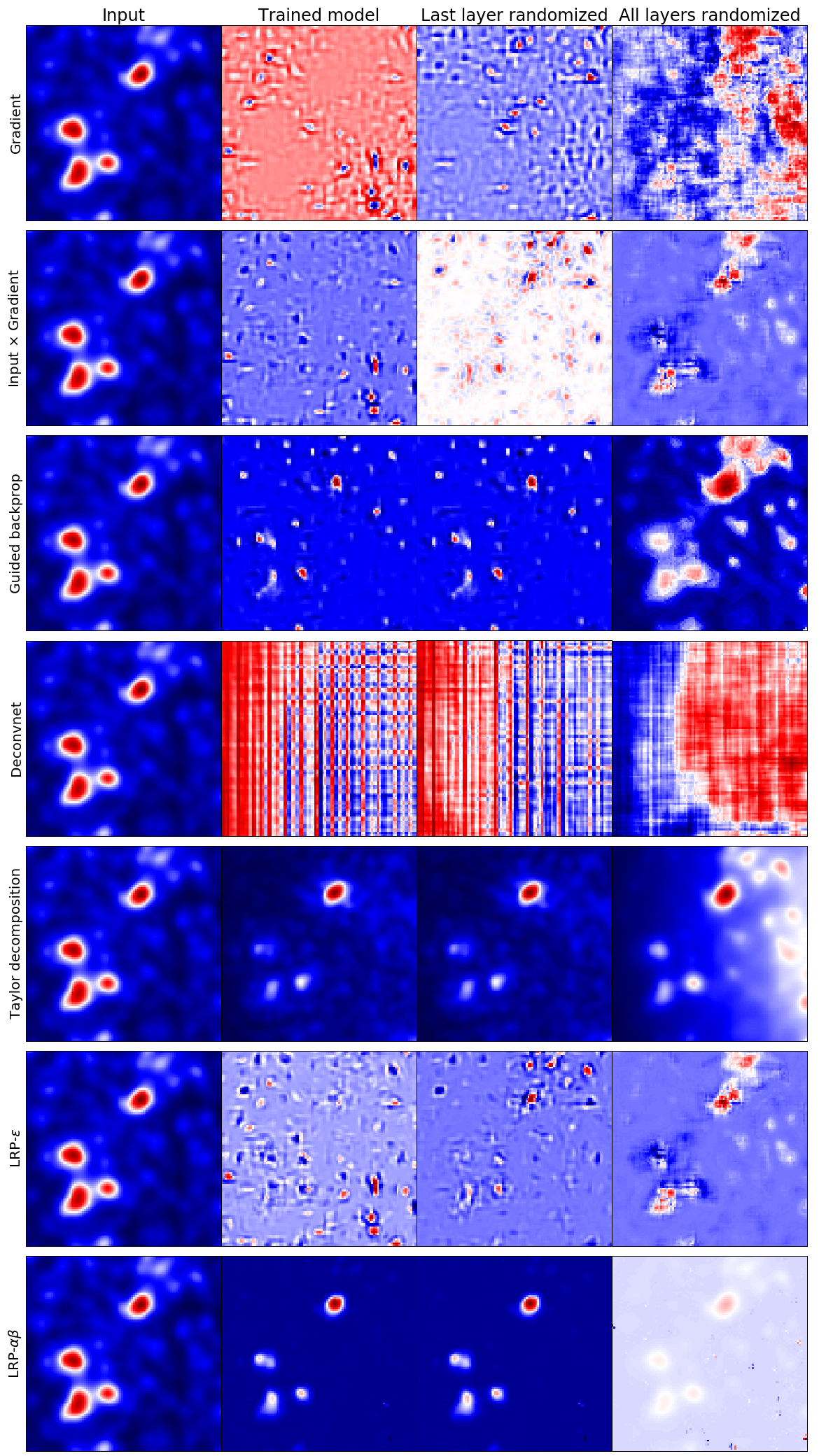}
    \caption{Examples of saliency maps for the output neuron of the DNN that encodes the parameter $\Omega_m$. The left-most column (`'Input'') shows a small region (100$\times$100 pixels, or $0.68\times0.68\,\deg^2$) of a $3.5\times3.5\deg^2$ $\kappa$ map from the fiducial cosmology. The second column (`'Trained model'') shows the region of the saliency maps that corresponds to the region of the input map on the left. The third column (`'Last layer randomized'') shows the same saliency map as the second column, computed on the fully trained model after randomizing the weights of the last (output) layer. The right-most column (``All layers randomized'') shows the same saliency map as columns 2-3, computed on a model where all the weights are randomized. Each row corresponds to a different saliency method. The scales for each image are omitted for clarity, since they do not influence the conclusions.}
    \label{fig:cnninterp.saliencies}
\end{figure*}

We illustrate the different saliency techniques, applied to our WL maps, in Fig.~\ref{fig:cnninterp.saliencies}. For simplicity, we show only a small $0.68\times0.68\,\deg^2$ patch of the larger $3.5\times3.5\deg^2$ $\kappa$ map in the fiducial cosmology, but other regions and maps show the same characteristics.
Also, for simplicity, Fig.~\ref{fig:cnninterp.saliencies} is built using the output of the DNN neuron that encodes its predictions for $\Omega_m$, but we have found identical conclusions when using the neuron corresponding to $\sigma_8$. To compute the saliency maps, we used the publicly available Python library \textit{\textbf{iNNvestigate}}~\citep{Alber2018}.

The column in Fig.~\ref{fig:cnninterp.saliencies} labeled ``Trained model'' shows the result of applying the different saliency methods to the same input map (shown in the left-most column labeled ``Input''). Visual inspection shows that the saliency maps from different methods can be qualitatively very different. Three of the propagation-based methods ({\it Guided backpropagation}, {\it Deep Taylor decomposition} and {\it LRP-$\alpha\beta$}) show a clear correlation with structures in the input map, assigning high relevance to high $\kappa$ regions, such as those around lensing peaks. The two gradient-based methods exhibit a more subtle correlation in which high-$\kappa$ regions have relatively low relevance, and high-relevance peaks are instead associated with low-$\kappa$ regions around local minima of the input maps. The {\it LRP-$\epsilon$} map is very similar to the {\it Input$\times$gradient} map. We attribute this to the small $\epsilon$ used (in the limit of $\epsilon=0$ the two methods are equivalent for our network architecture). Finally, the {\it Deconvnet} map exhibits some checkerboard artifacts, likely induced by the deconvolution scheme used~\citep{Kim2019}, and little correlation with the input $\kappa$ field.

Clearly, the different saliency methods provide very different answers to the basic question of ``which input pixels are more relevant'' to the DNNs output.  It is therefore important to find a criterion to choose the method(s) most appropriate to interpret the model in the present context. Past work has shown that some saliency methods lack robustness~\citep{Ghorbani2017, Kindermans2017}, and could be inappropriate for our combination of data and model.  To assess the robustness of each method, we performed a model parameter randomization test, following the tests performed in~\cite{Adebayo2018}. For each method, we computed saliency maps not only on the trained DNN, but also on the models that result from randomizing the networks' parameters. We performed this randomization incrementally, starting with only the output layer, all the way to the first convolutional layer. Methods that yield saliency maps that are insensitive to these randomizations fail the test, as the structures in these saliency maps cannot then stem from features the DNN has learned during training.

As an illustration, the third and fourth columns of Fig.~\ref{fig:cnninterp.saliencies} (labeled ``Last layer randomized'' and ``All layers randomized'') show the saliency maps computed on the model after randomizing the weights of the output layer, and the weights of all the layers, respectively. Visually, the gradient-based methods (and {\it LRP-$\epsilon$}) are very sensitive to the model's parameters, while propagation-based methods exhibit strong correlations between the saliency map computed on the trained and the random model.

To quantify the similarity between the saliency map computed from the model and from the model with all the layers randomized, we computed three measurements of association between both maps: the Pearson's r, Spearman rank-order correlation coefficient, and Kendall's $\tau$, reported in Table~\ref{table:cnn.interp.saliencies} for the maps displayed in Fig.~\ref{fig:cnninterp.saliencies} (the results do not depend on which map is used for the analysis). For all three measurements, the null hypothesis is that there is no relationship (or correlation) between the two maps. The only methods for which the null hypothesis cannot be rejected at high significance for all the three tests are the gradient-based methods. This is consistent with a past analysis of {\it Guided backprop} and {\it Deconvnet}~\citep{Nie2018}. Thus, in the rest of this paper, we will use the {\it Gradient} and the {\it Input$\times$gradient} saliency maps to interpret the DNN.

\begin{table*}\label{table:cnn.interp.saliencies}
\begin{center}
\begin{tabular}{l c c c c c c}
\hline
\hline
& \multicolumn{6}{c}{Correlation measurement}\\
\cline{2-7}
& \multicolumn{2}{c}{Pearson} & \multicolumn{2}{c}{Spearman} & \multicolumn{2}{c}{Kendall}\\
\cline{2-3}\cline{4-5}\cline{6-7}
Saliency method & Coefficient & p-value & Coefficient & p-value & Coefficient & p-value\\
\hline
{\bf Gradient} & {\bf -0.002} & {\bf 0.205} & {\bf -0.001} & {\bf 0.527} & {\bf -0.001} & {\bf 0.578} \\
{\bf Input$\times$gradient} & {\bf -0.003} & {\bf 0.119} & {\bf -0.000} & {\bf 0.944} & {\bf 0.000} & {\bf 0.785}\\
Guided backpropagation & 0.152 & 0.000 & -0.025 & 0.000 & -0.018 & 0.000\\
Deconvnet & -0.731 & 0.000 & -0.675 & 0.000 & -0.505 & 0.000 \\
Deep Taylor decomposition & 0.882 & 0.000 & 0.989 & 0.000 & 0.912 & 0.000 \\
LRP-$\epsilon$ & -0.011 & 0.000 & -0.004 & 0.031 & -0.003 & 0.024 \\
LRT-$\alpha \beta$ & 0.463 & 0.000 & 0.173 & 0.000 & 0.115 & 0.000 \\
\hline
\hline
\end{tabular}
\caption{Correlation measurements between the saliency maps computed from the fitted DNN, and the same architecture with all the parameters randomized (see columns labelled `'Trained model'' and `'All layers randomized'' in Fig.~\ref{fig:cnninterp.saliencies}). For all three tests (Pearson, Spearman, and Kendall), the null hypotesis is that there is no relationship between the two saliency maps. P-values are two-sided. Only the two gradient-based methods (first two rows, in boldface) pass this null test.}
\end{center}
\end{table*}

\subsection{Mapping attributions back to physical space} \label{cnninterp.attributionmapping} 

\begin{figure*}
    \centering
    \includegraphics[width=\linewidth]{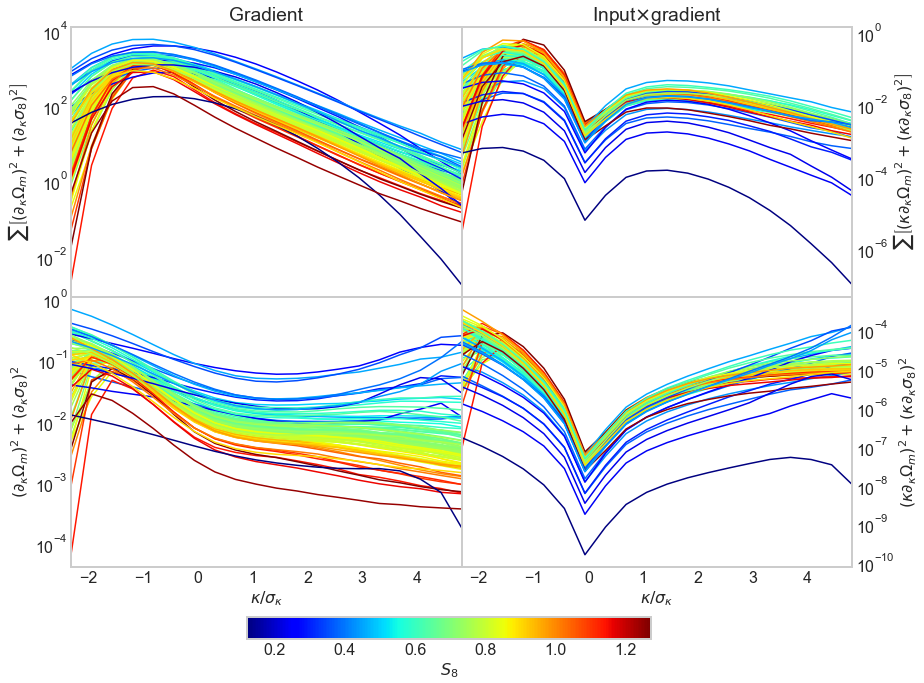}
    \caption{\textbf{Upper panels:} sum of the square of the pixel values in saliency maps as a function of $\kappa$ in the corresponding noiseless input maps. Each line is the test maps' average for one of the 101 cosmologies.
    \textbf{Lower panels:} same as upper panels, divided by the number of pixels in each $\kappa$ bin, giving the mean saliency$^2$ per pixel as a function of $\kappa$.
    Left panels correspond to saliency maps computed using the {\it Gradient} method, and right panels to saliency maps computed using the {\it Input}$\times${\it gradient} method. Each line is colored based on the value of $S_8=\sigma_8\left(\frac{\Omega_m}{0.3}\right)^{0.6}$.
    }
    \label{fig:cnninterp.kappa_attr}
\end{figure*}

\begin{figure*}
    \centering
    \includegraphics[width=\linewidth]{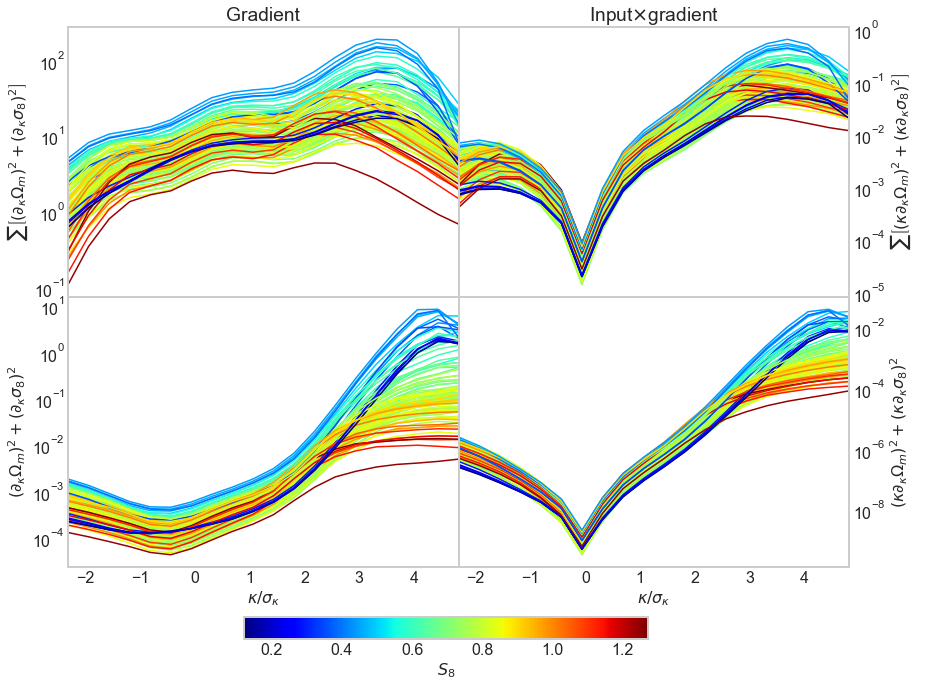}
    \caption{Same as Fig.~\ref{fig:cnninterp.kappa_attr} for convergence maps in the presence of shape noise.
    }
    \label{fig:cnninterp.kappa_attr.noisy}
\end{figure*}

We used the {\it Gradient} and {\it Input$\times$gradient} methods to analyze the distribution of the relevance for the DNN output, as a function of $\kappa$. For each saliency map in our data set, we measured sum of the square of the pixel values (to avoid cancellations by gradients or inputs of different sign) for the pixels within a given range of $\kappa$ values in the input map, and added the resulting squared saliencies for both $\Omega_m$ and $\sigma_8$. We selected 20 linear bins with $\kappa \in [-2.5, 5.0]$ in units of the r.m.s. $\kappa$ of each individual map. This measurement gives an estimate of the distribution of relevance in input space as a function of $\kappa$. We also measured the average relevance per pixel in each $\kappa$ bin. These two measurements, for each of the 101 cosmologies in the data set, are displayed in Fig.~\ref{fig:cnninterp.kappa_attr} for the noiseless case, and in Fig.~\ref{fig:cnninterp.kappa_attr.noisy} for the case with shape noise. The color of each line corresponds to the value of $S_8\equiv\sigma_8\left(\frac{\Omega_m}{0.3}\right)^{0.6}$ in each cosmology; this is approximately the best-measured combination, orthogonal to the direction of the degeneracy between the parameters $\Omega_m$ and $\sigma_8$.

With idealized, noiseless data, the most relevant pixels according to both gradient-based saliency methods, are those with extreme $\kappa$ values. Those at the negative tail of the $\kappa$ distribution are more relevant than those at the positive tail (see panels in the lower row of Fig.~\ref{fig:cnninterp.kappa_attr}). These pixels are rare, and the most relevant $\kappa$ regions are shifted towards the center of the $\kappa$ distribution (see panels in the upper row of Fig.~\ref{fig:cnninterp.kappa_attr}). Most of the relevance ends up being concentrated in regions with negative $\kappa$. These regions account for 86\% of the sum of the squared pixel values in the {\it Gradient} saliency maps, and 83\% in the {\it Input$\times$gradient} saliency maps. We note that the drop in relevance around $\kappa=0$ in the results from the {\it Input$\times$gradient} method is an artifact due to the zero value of the input.

In the presence of shape noise the relevance of negative-value pixels drops as they're dominated by noise, and that of extreme positive values increases. As a result, it is the high-$\kappa$ regions the ones that drive the predictions from the DNN (see upper panels in Fig.~\ref{fig:cnninterp.kappa_attr.noisy}). In this case negative $\kappa$ regions account for only 13\% of the sum of the squared pixels in the {\it Gradient} saliency maps, and 2\% in the {\it Input$\times$gradient} saliency maps. For comparison, regions with $\kappa > 3 \sigma_{\kappa}$ contribute 36\% and 68\%, respectively.

\section{Discussion and conclusions} \label{cnninterp.discussandconclude} 

In this study, we analyzed in detail a deep learning model that has been shown in previous work to learn cosmological parameters from simulated WL maps smoothed at an angular scale of $1\,{\rm arcmin}$.  Our aim was to understand which features in the simulated WL maps are used by the model to derive its predictions. First, we compared its performance with a suite of statistics commonly used in the WL community, individually and in combination, and evaluated the correlations between the DNN output, those statistics, and their linear combination that best fits the DNN output. Second, we borrowed a series of saliency methods from the field of image recognition and applied them to the DNN trained on simulated WL maps. We tested each method, and selected those that passed a null test of robustness, showing that they are sensitive to the learned weights of the DNN model, and are not directly derivable from the input maps. 
Finally, we used these methods to identify which pixels in simulated WL maps does the DNN use to discriminate between cosmological parameters. Our key findings are the following:

\begin{itemize}

    \item We generated a new suite of 100 simulations to accurately measure the covariance of the combination of five WL statistics (the power spectrum, lensing peaks, and three Minkowski functionals). We found that for noiseless, single-redshift simulated maps at $1\,{\rm arcmin}$ resolution, the third Minkowski functional, V$_2$ (representing the genus) is by far the most sensitive to cosmology, but it is also very sensitive to the presence of shape noise. Its constraining power for a galaxy density of $n_g=30\,{\rm arcmin^{-2}}$ is comparable to that of the other Minkowski functionals (and slightly worse than that from V$_1$). The relative performance of non-Gaussian statistics may differ for different datasets. For example, \cite{Zurcher2020} found that peaks outperform Minkowski functionals on simulated data smoothed with a Gaussian kernel with full-width-at-half-maximum of $10.5\,{\rm arcmin}$, a galaxy density of $5\,{\rm arcmin}^{-2}$ distributed among four tomographic bins, a different implementation of shape noise, and the inclusion of some systematics on $\kappa$ maps computed using the Born approximation on curved sky projections built on N-body simulations.
    
    \item The DNN can extract information not accessible through a combination of the power spectrum, lensing peaks, and Minkowski functionals. The addition of the DNN to those statistics reduces the credible region on the cosmological parameters of interest by $\approx$20\%, for both the noiseless and noisy cases.
    
    \item The DNN predictions are not highly correlated with the alternative statistics considered, nor can be reproduced using a linear combination of them. In the presence of shape noise, the correlation with statistics measured on high-$\kappa$ regions increases, but remains below 0.4 in absolute value.
    
    \item Saliency methods based on the back-propagation of the DNN output to input space were found to fail a simple robustness test: they are not sensitive to the values of the parameters that define the DNN. As a result, while (some) can provide attractive explanations in the form of attribution maps that highlight structures present in the input data, they do not represent which of these features are learned by the model. We urge practitioners to perform robustness tests when using these methods to interpret CNNs that use ReLUs as non-linearities.
    
    \item Gradient-based methods are sensitive to the parameters learned by the model, and as a result they are safe to use to interpret which features the DNN learns from the data. Another advantage of these methods is that their interpretation, for linear models, is straightforward: they correspond to regression coefficients or measure the contribution of each pixel to the output.
    
    \item Gradient-based methods show that for idealized, noiseless convergence maps, the most relevant pixels for the DNN are those with extreme values, at the tails of the $\kappa$ distribution. Negative $\kappa$ pixels are more relevant than positive $\kappa$ pixels, and when the number of pixels is taken into account, most of the relevance for the model output lies in regions with $\kappa<0$ (83-86\%). Shape noise dominates these low-$\kappa$ regions and suppresses their relevance. High-$\kappa$ regions dominate the relevance budget, so that $\kappa > 3 \sigma_{\kappa}$ pixels contribute with 36-68\% of the relevance.
    
\end{itemize}

While our result needs to be verified under more realistic treatments that include realistic galaxy distributions and noise, as well as systematic errors, they suggest that DNNs will be an attractive cosmological probe which primarily extract their information from relatively high-$\kappa$ regions. We find, however, that if shape noise can be mitigated, the DNNs derive a significant fraction of their cosmological sensitivity from negative $\kappa$ regions. 
This would have implications for the analysis of large future WL datasets.  Large voids, accounting for de-magnified and under-dense regions, have previously been found to contain most of the cosmological information in simulated maps with a galaxy redshift distribution and shape noise levels somewhat lower than considered here and appropriate for LSST~\citep{Coulton2020}.  These regions have also been shown to be less affected by baryonic physics, which are hard to capture accurately in simulations of growth of structure~\citep{Paillas2017, Coulton2020}. On the other hand, these regions have been shown to be sensitive to neutrino physics and modified gravity theories~\cite{Davies2018, Paillas2019, Kreisch2019}.

\section{Acknowledgements} \label{cnninterp.acknowledgements} 
We thank the anonymous referees, Dominik Z\"urcher, and Tomasz Kacprzak for useful comments that have resulted in an improved manuscript. We gratefully acknowledge support from NASA ATP grant 80NSSC18K1093, a JP Morgan Faculty Award, and the use of NASA's Pleiades cluster and the NSF XSEDE facility for simulations and data analysis in this study.

\bibliography{refs.bib}

\appendix

\section{Interpolation error on DNN emulator}\label{cnninterp.appendix_A}

Here we verify the robustness of our conclusions to possible interpolation errors of the emulator for the DNN. The output of the network as a function of the true cosmological parameters of input maps can be seen in Figures 2 and 4 of~\citep{Ribli2019b}.

First, we evaluate the interpolation error on each of the 101 cosmologies. To do so, for each cosmology we build an emulator using the test maps except those that correspond to that cosmology. Then, we compare the cosmological parameters for that cosmology with the corresponding output from the emulator. For the noiseless case, the interpolation error is at most 1.89\% for $\sigma_8$ at one of the boundaries of the sampled parameters, and well within 1\% in the neighborhood of the fiducial for both $\Omega_m$ and $\sigma_8$. For noisy data, the bias of the network's output in sparsely populated regions of parameter space results in larger interpolation errors, that can reach $\approx 28\%$ at the boundaries of the sampled parameter space, but remain at $\approx1\%$ in the vicinity of the fiducial cosmology. These results are consistent with those found for emulators of the convergence power spectrum and lensing peaks built from simulated data~\citep{Liu2015}.

\begin{figure*}
    \centering
    \includegraphics[width=\linewidth]{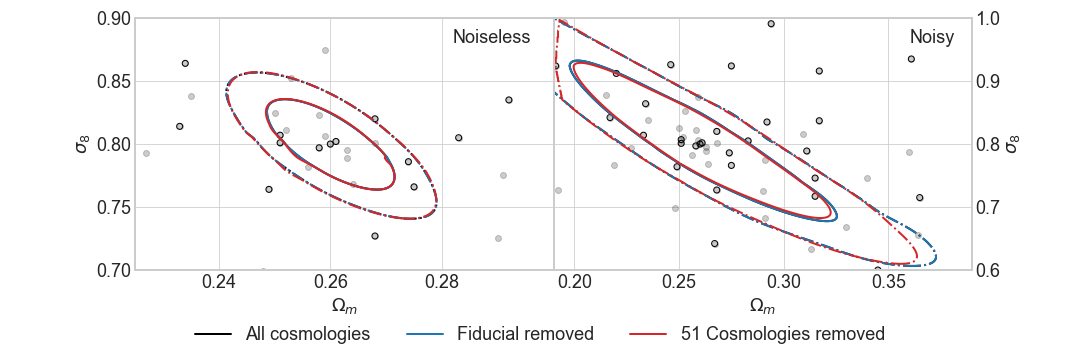}
    \caption{
    Comparison of the credible contours derived from the DNN and shown in Fig.~\ref{fig:cnn.interp.performance} and Fig.~\ref{fig:cnn.interp.performance.noisy}, with those derived using a DNN emulator built without data from the fiducial cosmology, and an emulator built using 50 random cosmologies from the original 101. The left panel shows the contours for noiseless data, and the right one for noisy data. The differences between the contours derived droping the fiducial and using all the cosmologies are smaller than the thickness of the lines. Dropping half the models (using 50, indicated by the outlined circles) has a minor effect.
    }
    \label{fig:cnn.interp.appendix_A}
\end{figure*}

Second, we estimate how the interpolation error propagates into the inferred parameters. We compute the credible contours using the emulator built excluding fiducial maps, and compare them to the contours derived using all the test maps (shown on Fig.~\ref{fig:cnn.interp.performance} and Fig.~\ref{fig:cnn.interp.performance.noisy}). Both sets of contours are indistinguishable from each other, as can be seen on Fig.~\ref{fig:cnn.interp.appendix_A}. Even when half the models are dropped when building the emulator, the resulting contours are just slightly perturbed. As a result, we do not expect our conclusions to be affected by interpolation errors in the emulator.

\end{document}